\documentclass[iop]{emulateapj}

\usepackage{verbatim}

\newcommand{\msun}{$M_{\odot}$}
\newcommand{\massmw}{$M=10^{10.5}~M_{\odot}$}
\newcommand{\sfr}{$M_{\odot}$~yr$^{-1}$}

\begin{document}

\title{The Structural Evolution of Milky Way-like Star Forming Galaxies since $\lowercase{z} \sim 1.3$\altaffilmark{*}}

\author{Shannon G. Patel$^1$, Mattia Fumagalli$^1$, Marijn Franx$^1$, Pieter G. van Dokkum$^2$, Arjen van der Wel$^3$, Joel Leja$^2$, Ivo Labb\'{e}$^1$, Gabriel Brammer$^4$, Rosalind E. Skelton$^2$, Ivelina Momcheva$^2$, Katherine E. Whitaker$^5$, Britt Lundgren$^6$, Adam Muzzin$^1$, Ryan F. Quadri$^7$, Erica June Nelson$^2$, David A. Wake$^6$, Hans-Walter Rix$^3$}

\altaffiltext{1}{Leiden Observatory, Leiden University, P.O. Box 9513, NL-2300 AA Leiden, Netherlands; patel@strw.leidenuniv.nl}
\altaffiltext{2}{Department of Astronomy, Yale University, New Haven, CT 06511, USA}
\altaffiltext{3}{Max-Planck-Institut fur extraterrestrische Physik, Giessenbachstrasse, D-85748 Garching, Germany}
\altaffiltext{4}{European Southern Observatory, Alonson de Cordova 3107, Casilla 19001, Vitacura, Santiago, Chile}
\altaffiltext{5}{Astrophysics Science Division, Goddard Space Center, Greenbelt, MD 20771, USA}
\altaffiltext{6}{Department of Astronomy, University of Wisconsin-Madison, Madison,WI 53706, USA}
\altaffiltext{7}{Observatories of the Carnegie Institution of Washington, Pasadena, CA 91101, USA}

\submitted{Accepted for publication in ApJ}

\altaffiltext{*}{Based on observations made with the NASA/ESA Hubble Space Telescope, obtained at the Space Telescope Science Institute. STScI is operated by the Association of Universities for Research in Astronomy, Inc. under NASA contract NAS 5-26555.}

\begin{abstract}
We follow the structural evolution of star forming galaxies (SFGs) like the Milky Way by selecting progenitors to $z\sim1.3$ based on the stellar mass growth inferred from the evolution of the star forming sequence.  We select our sample from the 3D-HST survey, which utilizes spectroscopy from the {\em HST} WFC3 G141 near-IR grism and enables precise redshift measurements for our sample of SFGs.  Structural properties are obtained from S\'{e}rsic profile fits to CANDELS WFC3 imaging.  The progenitors of $z=0$ SFGs with stellar mass \massmw\ are typically half as massive at $z\sim1$.  This late-time stellar mass growth is consistent with recent studies that employ abundance matching techniques.  The descendant SFGs at $z\sim0$ have grown in half-light radius by a factor of $\sim1.4$ since $z\sim1$.  The half-light radius grows with stellar mass as $r_e\propto M^{0.29}$.  While most of the stellar mass is clearly assembling at large radii, the mass surface density profiles reveal ongoing mass growth also in the central regions where bulges and pseudobulges are common features in present day late-type galaxies.  Some portion of this growth in the central regions is due to star formation as recent observations of H$\alpha$ maps for SFGs at $z\sim1$ are found to be extended but centrally peaked.  Connecting our lookback study with galactic archeology, we find the stellar mass surface density at $R=8$~kpc to have increased by a factor of $\sim2$ since $z\sim1$, in good agreement with measurements derived for the solar neighborhood of the Milky Way.  
\end{abstract}

\section{Introduction}

The structural formation of late-type star forming galaxies (SFGs) like the Milky Way has been studied through several complementary approaches.  Detailed studies of the ages, metallicities, and kinematics of stellar populations within the Milky Way itself afford a unique vantage point for viewing the assembly of such late-type galaxies -- although such an approach is limited in sample size and completeness \citep[see, e.g., ][]{rix2013}.  On the theoretical front, the formation of Milky Way-like SFGs has posed a challenge for simulations that aim to produce a structurally realistic analog at $z\sim0$ with a disk and an embedded bulge or pseudobulge \citep{scannapieco2012}.  This largely reflects the difficulty of modeling the ``sub-grid'' physical processes that impact the baryons \citep[although see, e.g.,][]{guedes2011b,stinson2013}.  Lookback studies provide another window into the formation of these galaxies.  Though several works have studied the structural properties of late-type galaxies in the distant universe, including the size-mass relation to $z\sim1$ and beyond \citep[e.g.,][]{barden2005,trujillo2006b}, the connection to the structural evolution of a typical disk galaxy, as it grows in mass, remains largely unexplored \citep[for comparisons to models, see][]{somerville2008,dutton2011}.

In this paper, we systematically select progenitors of Milky Way-like SFGs using the stellar mass growth inferred from the evolution of the star forming sequence and analyze their structural evolution.  We use {\em HST} WFC3 near-IR imaging that probes redder rest-frame wavelengths than most previous works and is therefore less sensitive to light from young stars.  Our selection provides a quantitative view for stellar mass build up in different radial regimes.  The method for computing the mass growth of SFGs has been discussed in detail by \citet{leitner2012} and complements other studies that connect progenitors and descendants using number densities \citep{vandokkum2010,patel2013}.  Given the distinct formation history of SFGs from quiescent galaxies (QGs), we use the former method here as it directly traces the star forming progenitors of galaxies like the Milky Way. 
In \citet{vandokkum2013}, we use the number density approach to study progenitors of galaxies of all types with the mass of the Milky Way.

We assume a cosmology with $H_0=70$~km~s$^{-1}$~Mpc$^{-1}$, $\Omega_M=0.30$, and $\Omega_{\Lambda}=0.70$.  Stellar masses are based on a Chabrier IMF \citep{chabrier2003}.  Magnitudes are in the AB system.

\section{Data and Analysis}

\begin{figure*}
\epsscale{1.2}
\plotone{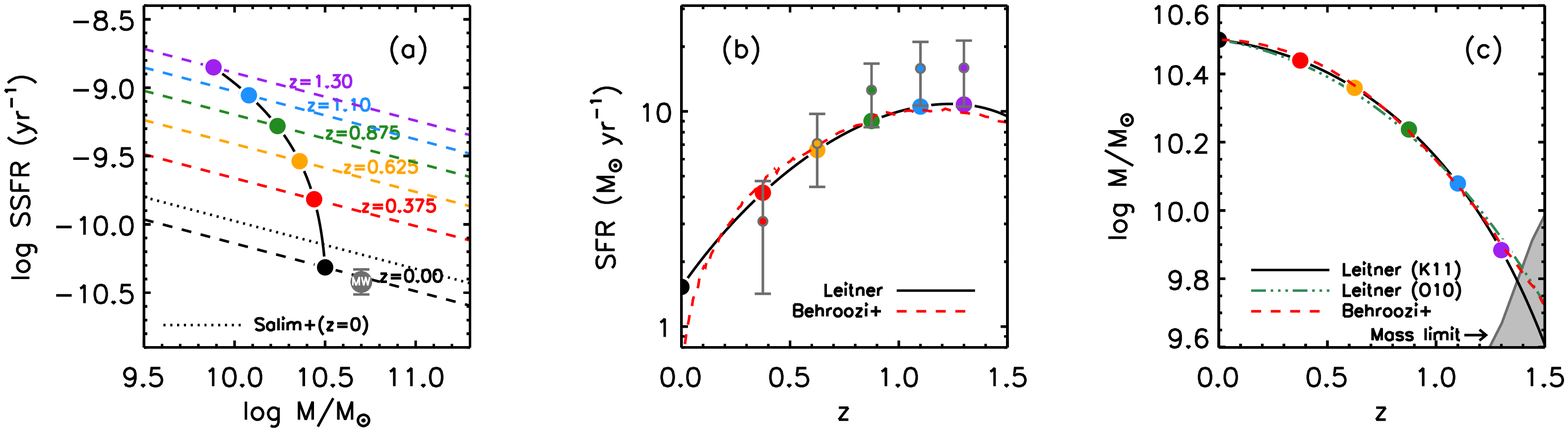}
\caption{(a) Evolution of SFGs in the SSFR-mass plane with final mass at $z\sim0$ of \massmw\ as computed by \citet{leitner2012} using the SSFR-mass relations at different redshifts (dashed lines) from \citet{karim2011}.  At a given redshift, new stellar mass is added according to the position of the SFG on the star forming sequence and mass loss from stellar evolution is also accounted for.  The slight offset between the extrapolated $z=0$ relation from that of \citet{salim2007} does not impact our results (see text).  The Milky Way (gray circle) lies within the observational scatter of the $z=0$ relation.  (b) SFH of the SFGs tracked in panel (a).  The data points with error bars represent IR+UV SFRs for a subset of our sample with deep MIPS: there is good general agreement with the radio based SFRs from \citet{karim2011}. (c) Mass growth history of the SFGs tracked in panel (a).  We select galaxies at the indicated masses and redshifts in this work (i.e., colored points on the black curve).  Since $z=1$, the SFGs grow in stellar mass by a factor of $\sim2$.  Mass growth histories derived from alternative SFR measurements give very similar results (e.g., green curve, which uses far-infrared SFRs from \citet{oliver2010}).  For comparison, the SFH and mass growth for galaxies with the same final mass from the abundance matching work of \citet{behroozi2012} is shown by the dashed red line.  There is good agreement between the two different methods.} \label{fig_selection}
\end{figure*}

We employ data from the 3D-HST survey ($v2.1$) to carry out our analysis.  The observations and data reduction procedures are explained in detail in \citet{brammer2012} and Skelton~et~al. (in prep.).  The {\em HST} WFC3 G141 near-IR grism observations are the centerpiece of 3D-HST and cover the CANDELS fields.  In this work, we use the three fields with currently available WFC3 based structural parameters from the literature: COSMOS, GOODS-S, and UDS.  Grism redshifts were measured using a modified version of EAZY \citep{brammer2008}.  The procedure fits template SEDs to both the photometry and grism spectroscopy, enabling precise redshift measurements \citep[][]{brammer2012} for our sample of SFGs at high redshift.  Objects selected in this work had full spectral coverage and less than $<50\%$ integrated spectral contamination.

Stellar masses were measured with FAST \citep{kriek2009}.  We determine the stellar mass limit at our highest redshifts ($z=1.4$) to be $M\sim 10^{9.8}$~\msun\ (see Figure~\ref{fig_selection}(c)) using a similar technique to that of \citet{marchesini2009}.  This conservative limit accounts for galaxies with high $M/L$ such as reddened SFGs.  Rest-frame $U-V$ and $V-J$ colors were measured with EAZY in order to distinguish SFGs from QGs (Section~\ref{sec_uvj}).

Structural properties were obtained from \citet{vanderwel2012} who used GALFIT \citep{pengc2010} to fit S\'{e}rsic profiles to the CANDELS {\em HST} WFC3 $J_{125}$ and $H_{160}$ imaging \citep{grogin2011,koekemoer2011}.  Only objects with no flags were selected.  The effective radii reported here are circularized.  The $J_{125}$ imaging is used at $0.25<z<1$ while the $H_{160}$ imaging is used at $1<z<1.4$, with our median galaxy sampling rest-frame $\sim7100$~\AA.  The PSF FWHM/2 of the $H_{160}$ imaging is $\sim 0\farcs09$, corresponding to a physical radius at our highest redshifts of $\sim 0.8$~kpc.

We employ an SDSS reference sample at $z=0.05$ using stellar masses and SFRs from the MPA/JHU catalogs \citep{brinchmann2004} and $i$-band S\'{e}rsic fits from the NYU-VAGC \citep{blanton2003d,blanton2005}.

$ $

\section{Selection}

\subsection{Stellar Mass Growth History from an Evolving Star Forming Sequence}

\begin{figure*}
\epsscale{1.2}
\plotone{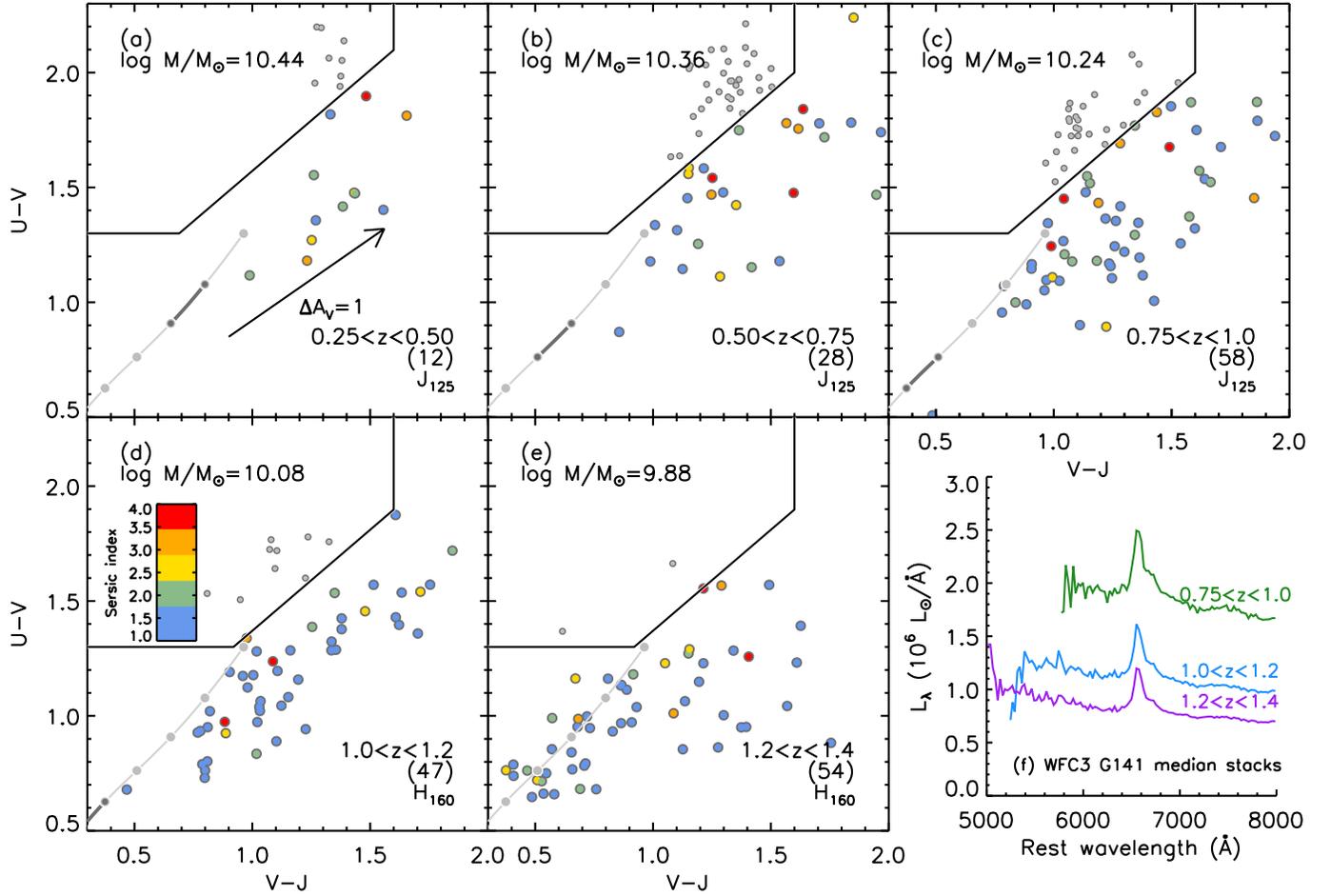}
\caption{(a)-(e) Rest-frame $U-V$ versus $V-J$ at different redshifts, with galaxies selected to have the progenitor mass ($\pm 0.1$~dex) from Figure~\ref{fig_selection}(c).  The \citet{williams2009} boundary separating QGs from SFGs (bottom right) is used to select the latter.  The gray curve indicates the evolution in $UVJ$ colors for the SFH shown in Figure~\ref{fig_selection}(b) with the darker segment representing the colors spanning the given redshift interval.  The SFGs are color-coded according to their S\'{e}rsic indices.  Sample sizes are indicated in the bottom right. (f) WFC3 G141 median stacked spectra of SFGs for redshift bins with observable H$\alpha$.} \label{fig_uvj}
\end{figure*}

\begin{figure*}
\epsscale{1.2}
\plotone{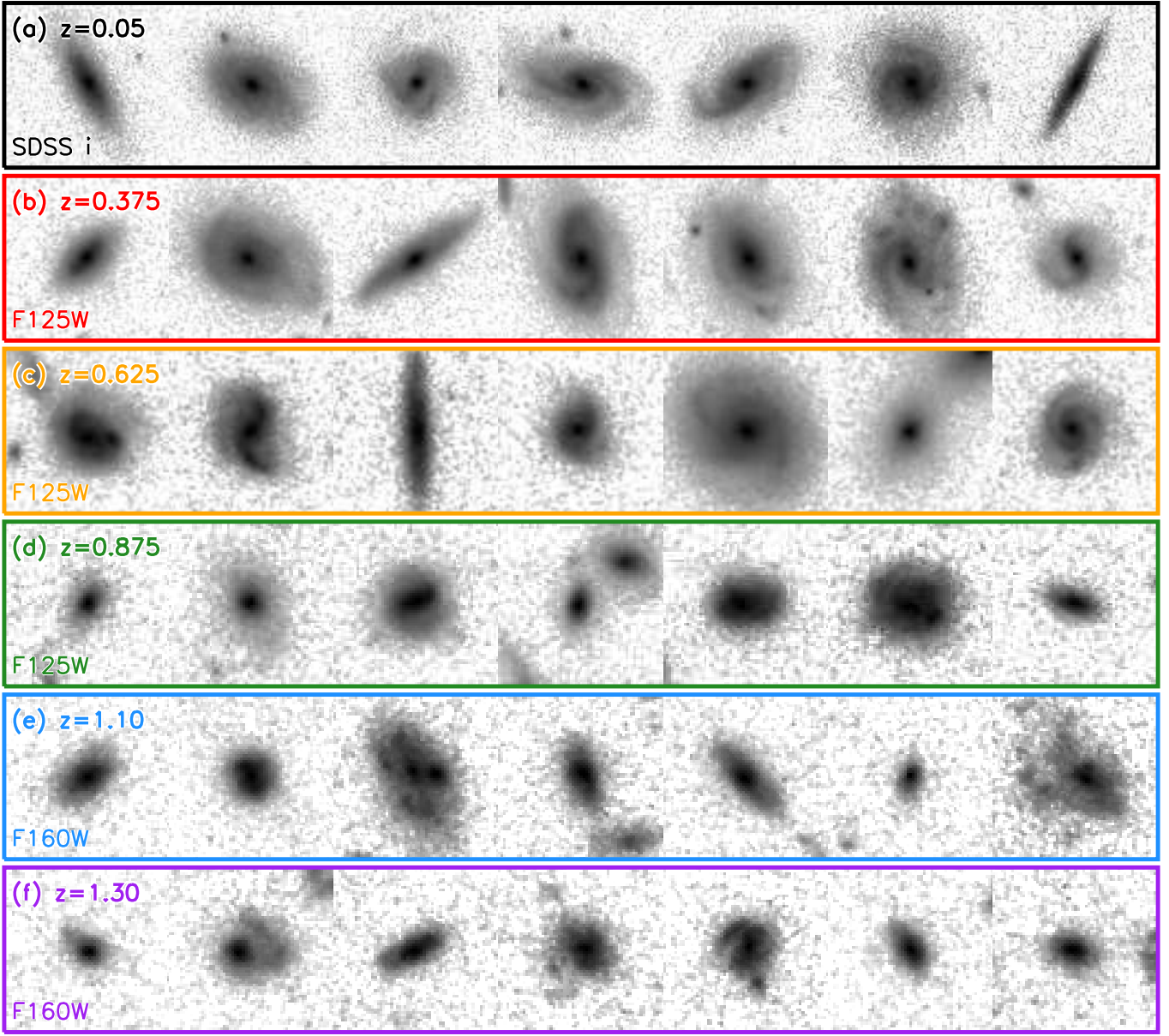}
\caption{Random SDSS $i$-band and {\em HST} WFC3 postage stamps for progenitors of SFGs with a final mass at $z=0$ of \massmw.  Each stamp is $\sim 30$~kpc on a side.  The median stellar mass decreases to high redshift according to Figure~\ref{fig_selection}(c) where the sizes appear smaller.} \label{fig_mwps}
\end{figure*}

\begin{figure*}
\epsscale{1.15}
\plottwo {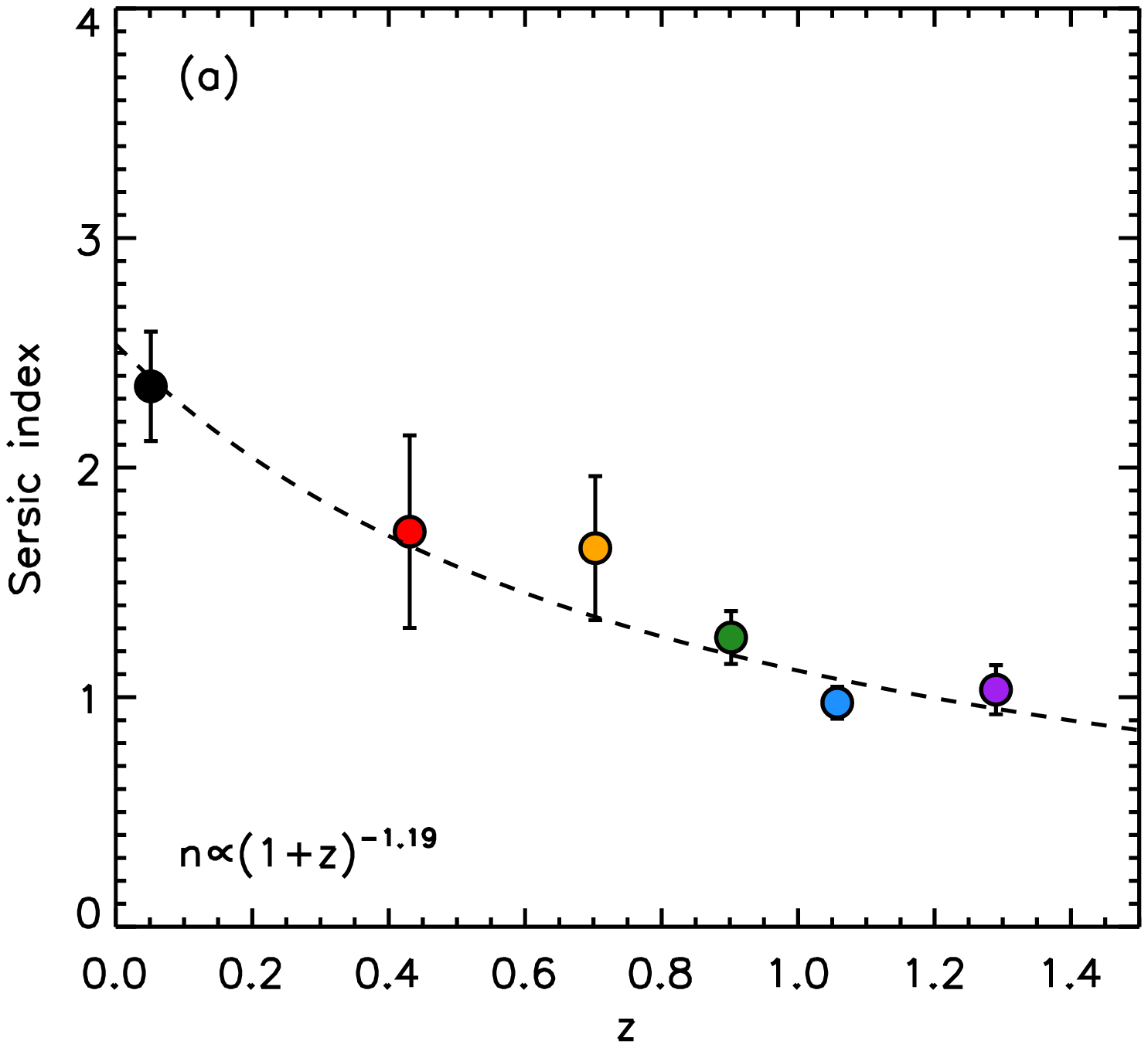}{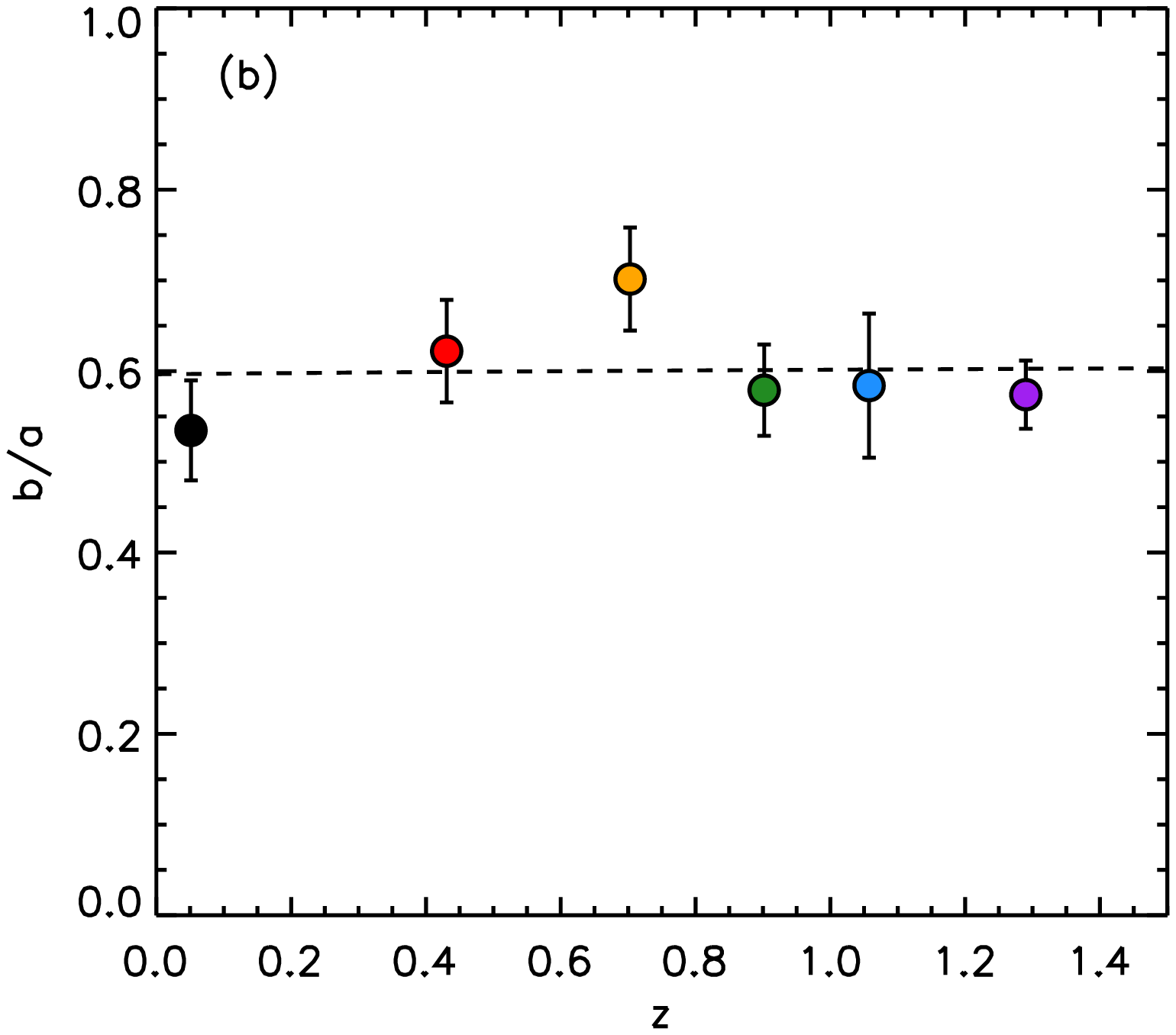}
\plottwo {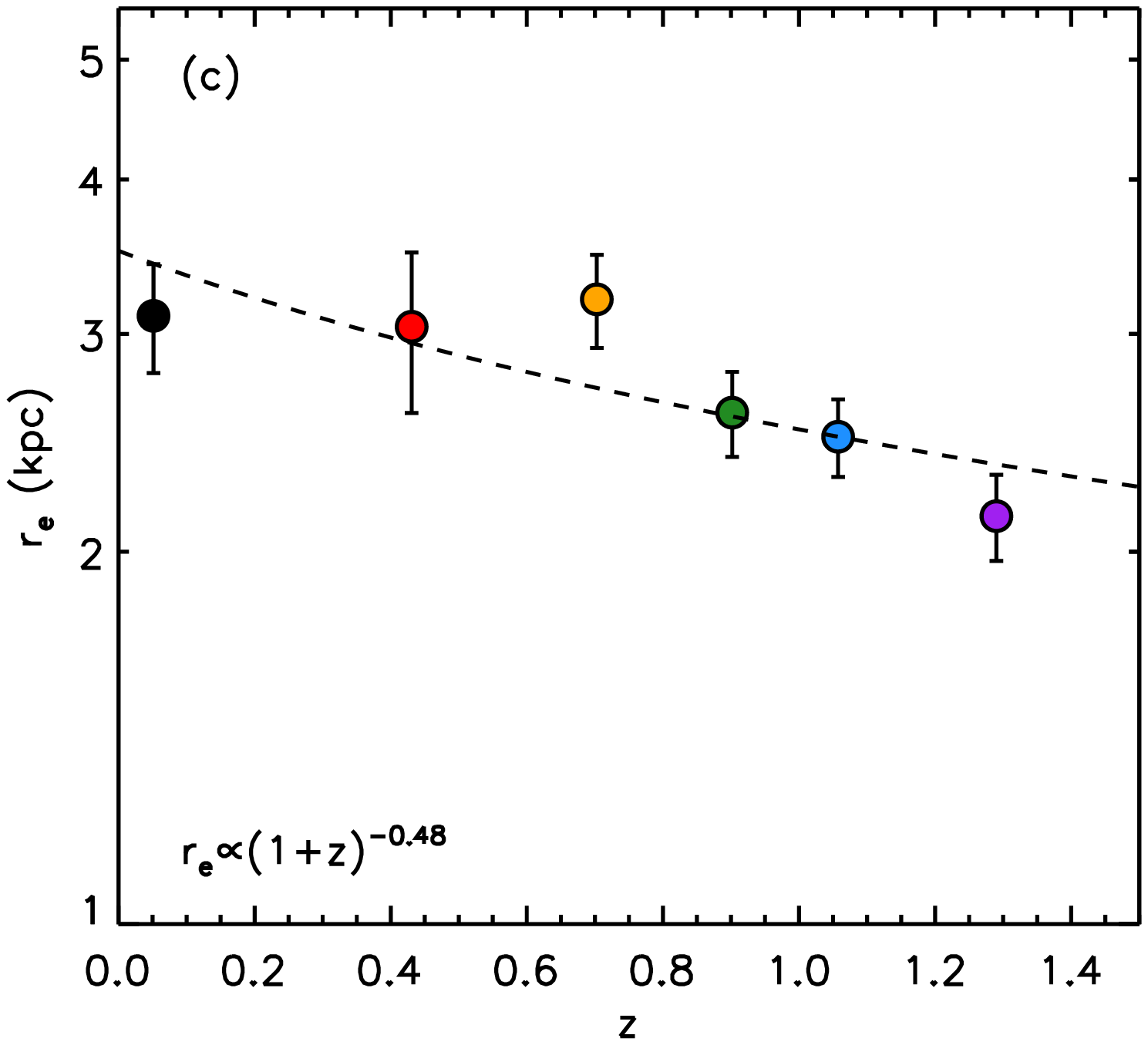}{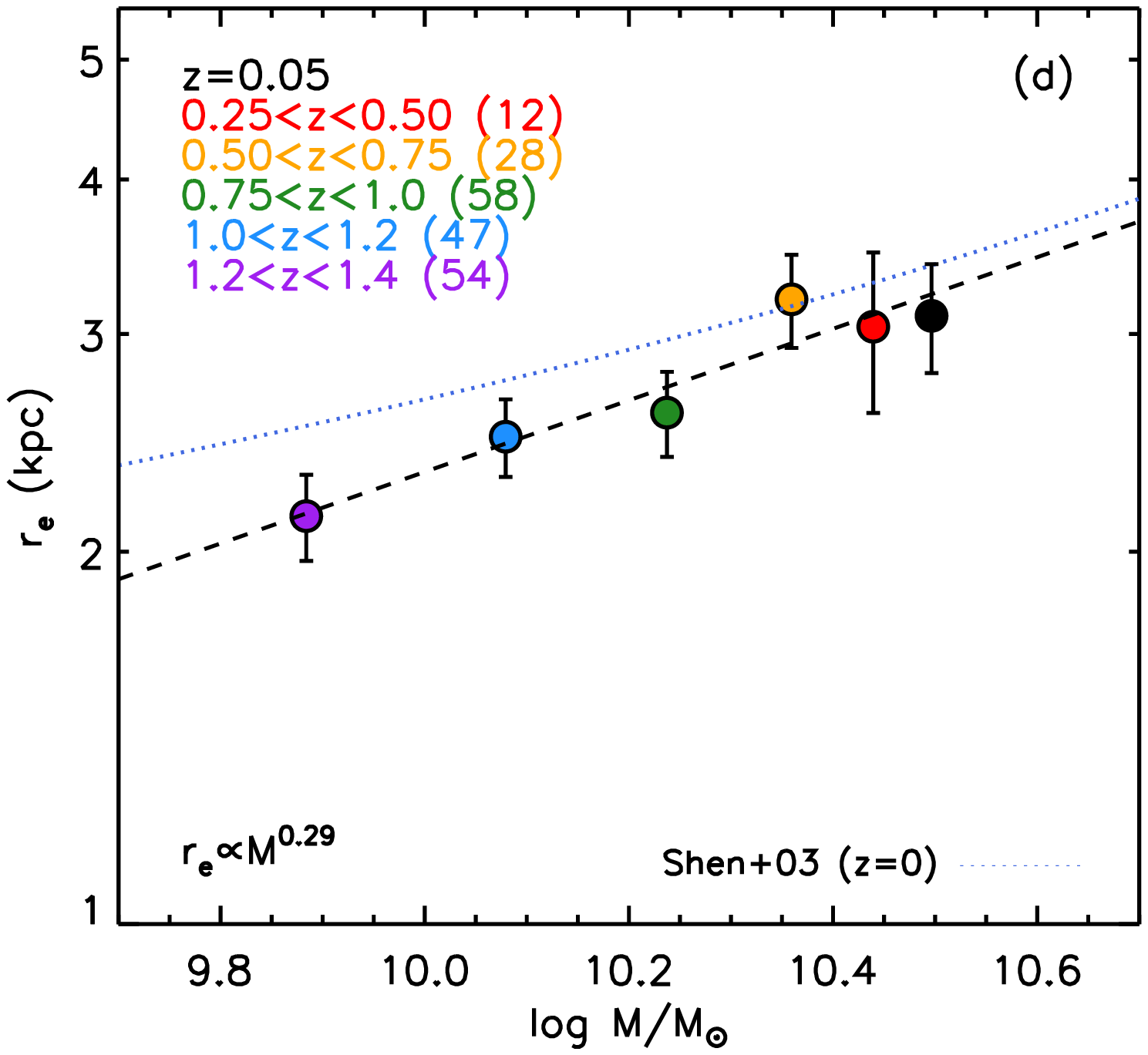}
\caption{(a) Median S\'{e}rsic index versus redshift for progenitor SFGs. (b) Median axis ratio versus redshift. (c) Median half-light radius versus redshift.  The error bars represent the bootstrapped uncertainty on the median.  Fits of the form $(1+z)^{\alpha}$ are indicated for the S\'{e}rsic indices and half-light radii, both of which increase toward low redshift.  (d) Evolution of $r_e$ in the size-mass plane.  The dotted blue line is the $z\sim0$ SDSS size-mass relation ($z$-band) for late-type galaxies from \citet{shen2003}.  The dashed line is a fit to the data of the form $r_e\propto M^{\alpha}$, where $\alpha=0.29\pm0.08$.  Since $z=1$, the half-light radii of the SFGs have grown by a factor of $\sim 1.4$.} \label{fig_structure}
\end{figure*}

\begin{figure*}
\epsscale{1.15}
\plottwo{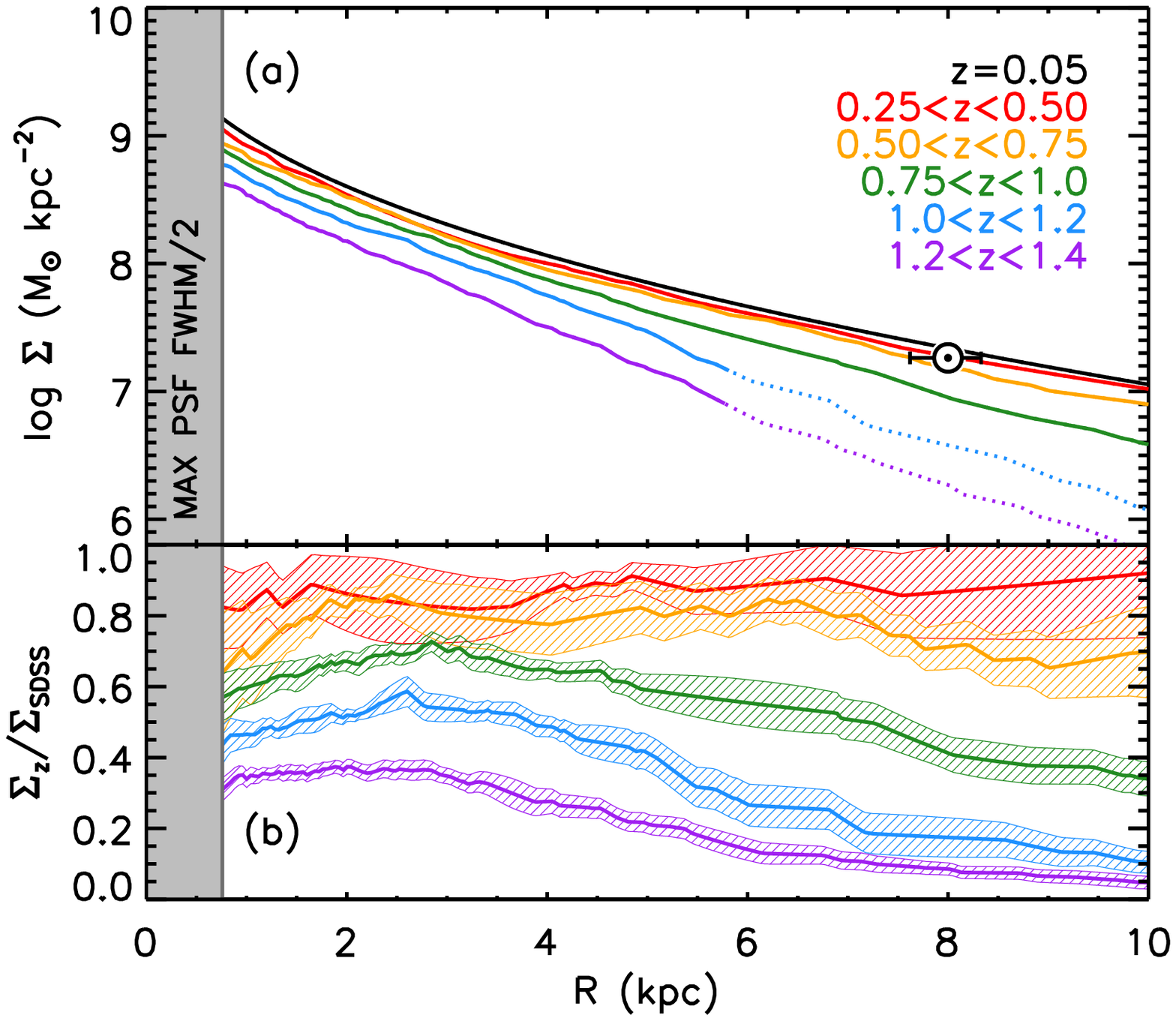}{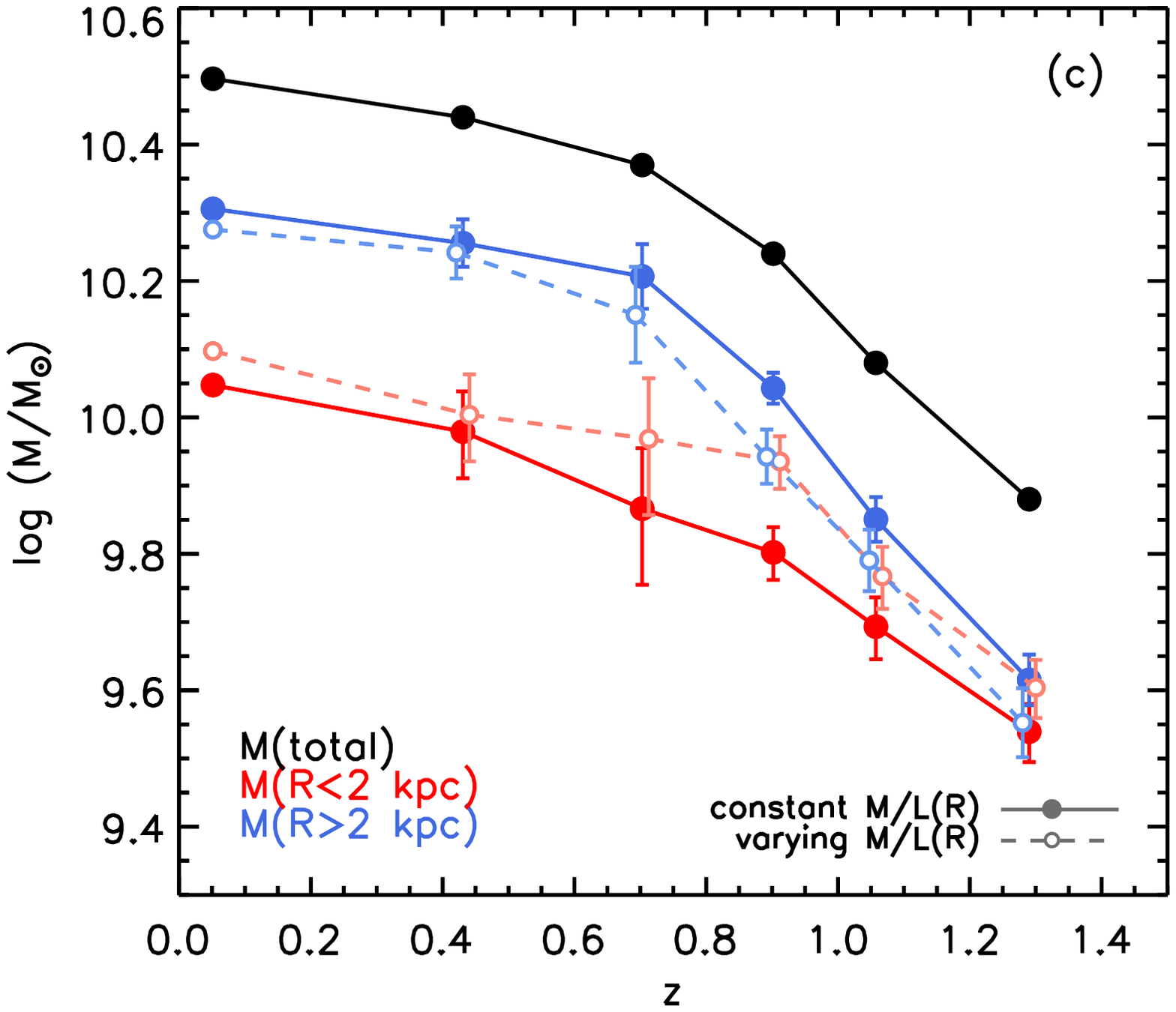}
\caption{(a) Median combined mass surface density profiles.  The dotted portion of each curve indicates where the uncertainty becomes larger than $>20\%$.  The shaded gray region indicates the maximum WFC3 PSF FWHM/2.  The solar symbol indicates the mass surface density in the solar neighborhood as computed by \citet{bovy2012b} but scaled down by $0.2$~dex to account for the difference in mass between the Milky Way and our $z\sim0$ descendants.  (b) The ratio of the mass profiles at higher redshifts to the SDSS mass profile.  Shaded region indicates the uncertainty.  The factor of $2.4\pm0.3$ mass growth from $z\sim0.9$ to $z\sim0$ at $R_0=8$~kpc agrees well with measurements in the solar neighborhood from \citet{aumer2009}.  (c) Median stellar mass growth for the central ($R<2$~kpc, red) and outer regions ($R>2$~kpc, blue).  The total stellar mass growth is shown in black.  While most of the new mass growth since $z\sim1$ has taken place in the outer regions, mass has also been added in the central regions of these SFGs where bulges and pseudobulges are common features.  Some portion of the growth in the central regions is due to star formation as H$\alpha$ maps for SFGs at $z\sim1$ are centrally peaked \citep{nelson2013}.  Dashed lines with open circles indicate the mass growth when accounting for $M/L$ differences in the two radial regimes (see text).} \label{fig_mass_profiles}
\end{figure*}

In this work, we trace the formation of SFGs with a final mass of \massmw, which is slightly below ($\sim0.2$~dex) that of the Milky Way, so that we can compare various properties to galaxies of a similar final mass from other works that employ different progenitor-descendant linking methods \citep[e.g.,][]{behroozi2012}.  We note however that we arrive at similar qualitative conclusions when tracing progenitors of Milky Way-mass SFGs (i.e., $M_{z=0}\sim10^{10.7}$~\msun).

Selecting progenitors of $z\sim0$ SFGs requires knowledge of their mass growth history so that one can select galaxies of the proper progenitor mass at a given redshift.  For galaxies that assemble most of their stars from in-situ star formation, one can infer the mass growth from the evolution in the SFR-mass relation (or SSFR-mass).  The Milky Way, with a stellar mass of $M\sim5\times10^{10}$~\msun\ \citep{hammer2007} and a SFR of $1.9\pm0.4$~\sfr\ \citep{chomiuk2011}, falls within the observational scatter of the SSFR-mass relation of \citet[][their Eq.~11, corrected for evolution to $z=0$ assuming Equation~\ref{eq_ssfr} here; see Figure~\ref{fig_selection}(a)]{salim2007}.  This star-forming sequence has been extensively studied to high redshifts \citep[e.g.,][]{noeske2007,karim2011,whitaker2012b,fumagalli2012}.  The method for determining the mass growth from this relation is simple: for a given redshift interval, new stellar mass is added to the existing mass based on the location of a galaxy in the SSFR-mass plane and mass loss from stellar evolution is accounted for from simple stellar population models.  The assumption that most nearby SFGs were star forming at high redshift is supported by the small scatter about the SSFR-mass relation as well as by the vastly different structural properties between SFGs and QGs \citep[e.g.,][]{franx2008}.

In this work, we use the mass growth computed by \citet{leitner2012}, who derived it from the SSFR-mass relations of \citet{karim2011}.  Figure~\ref{fig_selection}(a) shows SSFR-mass relations at different redshifts (dashed lines) measured by \citet{karim2011} from 1.4 GHz radio stacking.  They employ the parameterization: 

\begin{eqnarray}
SSFR~\propto M^{\beta} (1+z)^n  \label{eq_ssfr}
\end{eqnarray}
where $\beta=-0.35$ and $n=3.45$.  Though the $z=0$ relation based on Equation~\ref{eq_ssfr} is an extrapolation of the \citet{karim2011} data, its slight difference from lower redshift observations \citep[e.g.,][]{salim2007} does not significantly impact the mass growth at these late times.  The track shown in Figure~\ref{fig_selection}(a) (black curve) indicates the trajectory in the SSFR-mass plane for the SFGs selected for study in this work with final mass at $z=0$ of \massmw.  The corresponding SFH is shown in panel (b) and the mass growth with redshift in panel (c).  The points with error bars (including a $30\%$ systematic uncertainty) in panel (b) indicate the median SFR \citep[IR+UV, computed similarly to][]{franx2008} of our sample in the GOODS fields (i.e., deep MIPS) for our $UVJ$-selected (see below) SFGs at the corresponding redshifts and masses from panel (c).  While SFR estimates can vary due to systematics between different tracers \citep[e.g.,][]{wuyts2011,leitner2012} there is good general agreement between our SFR measurements and those of \citet{karim2011}.  The green curve in panel (c) indicates the mass growth computed by \citet{leitner2012} when employing the far-infrared derived SFR data of \citet[][$\beta=-0.15$, $n=3.36$]{oliver2010}.  The small difference ($<0.05$~dex) compared to the adopted mass growth from the \citet{karim2011} SFR data (black curve) indicates that variations in the sample selection due to uncertainties in the mass growth from different SFR data are small and therefore does not impact our qualitative conclusions.  The dashed red curves in panels (b) and (c) show the corresponding SFH and mass growth for the same final mass galaxy from the abundance matching work of \citet{behroozi2012}.  There is good agreement between the two different methods.

The results above imply significant stellar mass growth in our SFGs below $z<1$ (factor of $\sim2.2$).  Late mass growth at $z<1$ is also found by other works \citep[e.g.,][]{yang2012,moster2013}.  SFHs of the Milky Way disk derived from stellar properties also suggest significant late time assembly \citep[e.g.,][]{rochapinto2000c,aumer2009}.  Finally, disks are more commonly formed in hydrodynamical simulations that favor late time assembly \citep{scannapieco2012}.

\subsection{Selection of Star Forming Progenitors to $\lowercase{z} \sim 1.3$} \label{sec_uvj}

Given the observational scatter about the SSFR-Mass relation, in practice we select all SFGs in a given redshift and mass bin.  Figure~\ref{fig_uvj} shows rest-frame $U-V$ vs. $V-J$ diagrams for galaxies in narrow mass bins ($\pm0.1$~dex) centered on the progenitor mass at a given redshift.  We use this $UVJ$ selection to identify SFGs, which occupy the bottom right portion of these diagrams \citep{williams2009,patel2012}.  The gray curve in the figure indicates the evolution in $UVJ$ colors for the dust-free SFH shown in Figure~\ref{fig_selection}(b) with the darker segment representing the colors spanning the given redshift interval.  These $UVJ$ colors are extended redward depending on the amount of reddening from dust (see, e.g., the reddening vector in the figure).  The descendants, which are more massive, become redder toward low redshift likely due to their aging stellar populations.  We note that the \citet{williams2009} boundary that is used to distinguish SFGs from QGs shifts to redder $U-V$ colors toward low redshift and therefore accounts for the general decline in galaxy SFHs.  In the SDSS sample, galaxies with SSFR~$>10^{-11}$~yr$^{-1}$ were selected as SFGs.  Figure~\ref{fig_mwps} shows random SDSS $i$-band and {\em HST} WFC3 postage stamps for progenitor SFGs at different redshifts.

$ $\\
$ $

\section{Results}

\subsection{Structural Evolution}

We examine the evolution in various structural parameters for our sample of progenitor SFGs in Figure~\ref{fig_structure}.  Panel (a) shows that most of the progenitors at high redshift have close to exponential profiles while those at lower redshift have slightly higher S\'{e}rsic indices.  The median axis ratio in panel (b) remains relatively constant at $b/a\approx0.60$, a low value which for a population of randomly inclined disks modeled as oblate spheroids would imply an intrinsic axis ratio of $\sim0.3$.  The low S\'{e}rsic indices and low axis ratios for the progenitor SFGs at higher redshifts is suggestive of disks.

Half-light radii provide a first order view into the distribution of stellar mass for galaxies in our sample.  Figure~\ref{fig_structure}(c) shows the evolution of the median half-light radius with redshift.  Though we follow the median of a given property, it is important to note that at a given redshift and stellar mass, SFGs display a diversity of property values.  The constant scatter with redshift about the size-mass relation (van der Wel~et~al., in prep), however, suggests that the increasing intercept of this relation toward low redshift is driven by evolution in the SFG population as a whole.  The black dashed line represents a fit to the data of the form:

\begin{eqnarray}
r_e=\beta(1+z)^{\alpha} \label{eq_re_redshift}
\end{eqnarray}
where $\beta=3.5\pm0.3$~kpc and $\alpha=-0.48\pm0.15$.  We note that all of the SDSS data points in Figure~\ref{fig_structure} include a $10\%$ error added in quadrature to account for systematics between different measurement methods \citep[e.g., Figure~A1 in][]{guo2009}.  Since $z=1$, the median half-light radius for these SFGs has grown by a factor of $\sim1.4$, indicative of some amount of inside-out growth.

Figure~\ref{fig_structure}(d) shows the evolution of the progenitors in the size-mass plane.  For reference, the SDSS $z\sim0$ size-mass relation for late-type galaxies from \citet{shen2003} is indicated by the dotted blue line.  The dashed line represents a fit to our sample of the form:

\begin{eqnarray}
r_e\propto M^{\alpha}
\end{eqnarray}
where $\alpha=0.29\pm0.08$.  Interestingly, the value of $\alpha$ measured here for SFGs is much smaller than that for QGs \citep[$\alpha_{\rm QG} \approx2$;][]{vandokkum2010,patel2013}.  SFGs generally appear to evolve near the local scaling relation below $z\lesssim1.3$, consistent with slow evolution in the intercept of the size-mass relation for SFGs (van der Wel~et~al., in prep).  Finally, we note that none of the results presented in Figure~\ref{fig_structure} change significantly when using the mass growth derived from the \citet{oliver2010} SFRs (green curve in Figure~\ref{fig_selection}(c)).

\subsection{Stellar Mass Growth in the Central and Outer Regions at $z<1.3$}

Stellar mass surface density profiles provide a more detailed look at the distribution of mass within galaxies.  Figure~\ref{fig_mass_profiles}(a) shows the median combined mass surface density profiles for our sample of SFG progenitors.  These profiles were constructed in a similar manner to that of \citet{patel2013}, where the best-fit single component S\'{e}rsic profiles were stacked to create the median light profile.  A single component S\'{e}rsic profile is generally found to be a good representation of the light profile for individual galaxies at high redshifts \citep{szomoru2012,mosleh2013}.  Prior to stacking, each light profile was converted into a mass profile by normalizing the light within $R=20$~kpc to the total stellar mass of each galaxy.  While a more thorough analysis would be required to properly account for $M/L$ gradients \citep[see, e.g.,][]{szomoru2013}, the present analysis provides a first order view into the evolution in the radial distribution of stellar mass.  Below, however, we also examine how different $M/L$ values in the central and outer radial regimes impact the measured stellar mass growth.  In addition, we will explore the impact of $M/L$ gradients for a broader sample in followup work.  Figure~\ref{fig_mass_profiles}(b) shows the ratio of the mass profiles at higher redshifts to the SDSS mass profile.  Below $z\lesssim 1.3$, stellar mass is continually built up at all radii.

Figure~\ref{fig_mass_profiles}(c) shows the growth in the projected mass inside (red) and outside (blue) of $R=2$~kpc (i.e., $\approx r_e$ at $z\sim1.3$).  Clearly more mass has assembled at larger radii since $z\sim1.3$, leading to larger $r_e$ toward low redshift.  However, the amount of stellar mass in the central regions, where bulges and pseudobulges are common structural features in nearby late-type galaxies \citep{weinzirl2009}, has also increased.  Some portion of the growth in the central regions is due to star formation as H$\alpha$ maps for SFGs at $z\sim 1$ are centrally peaked \citep{nelson2013}.  This is qualitatively consistent with recent observations of stars in the Milky Way bulge that display a wide range of metallicities and ages, implying an extended formation history \citep{bensby2013}.  The ongoing mass assembly in the central regions may point to secular processes \citep[e.g.,][]{kormendy2004} as an important channel for bulge growth for SFGs in the stellar mass regime studied here.

We estimate the impact on the stellar mass growth in the two different radial regimes due to variation in $M/L$ ratios.  These $M/L$ ratios can be derived from the correlation with galaxy colors \citep[e.g.,][]{bell2001}.  For each galaxy in our sample, we use the PSF deconvolved best-fit GALFIT models in the $J_{125}$ and $H_{160}$ bands to estimate the observed $J_{125} - H_{160}$ colors inside and outside of $R=2$~kpc ($R=10$~kpc was used for the upper bounds on the outer radial regime).  The median color difference between the different radial regimes (outer minus central) is, from low ($0.25<z<0.5$) to high redshift ($1.2<z<1.4$), $\Delta(J_{125} - H_{160})=-0.09,-0.11,-0.13,-0.10$, and $-0.09$~mag.  The outer regions are therefore bluer at all redshifts.  The $\Delta(J_{125} - H_{160})$ values were then converted into $\Delta(\log M/L_X)$ values using the slope of the correlation between $\log M/L_X$ and $J_{125} - H_{160}$,  where $X$ represents the observed $J_{125}$ (for the sample at $z<1$) and $H_{160}$ ($z>1$) bands.  In deriving these $M/L$-color correlations at different redshifts, we employed a broad sample of galaxies, including both QGs and SFGs above a mass of $M>10^{9.5}~M_{\odot}$ so that a wide array of stellar populations were sampled.  From low to high redshift, we find $\Delta(\log M/L_X)/\Delta(J_{125} - H_{160})=0.44,1.33,1.74,1.33,$ and $1.45$~dex~mag$^{-1}$.  The scatter in $\log M/L_X$ about these correlations is $0.15-0.19$~dex.  While some of this scatter is caused by working in the observed frame with a sample spanning a range of redshifts, it is similar to what is found in other recent works \citep[e.g., $0.13-0.28$~dex in][]{szomoru2013}.  We multiply our median light profile at a given redshift by the step function implied by $\Delta(\log M/L_X)$ (re-normalizing to the median mass of the sample) and integrate as before to determine the mass in the central and outer regions (open circles with dashed lines in Figure~\ref{fig_mass_profiles}(c)).  For the SDSS light profile, we apply the $i$-band $M/L$ gradient computed by \citet[][their Fig.~3]{tortora2011b}.  We adopt the average value between their ETG and LTG samples ($\Delta(\log M/L_i)/\Delta(\log R/R_e) \approx -0.1$) since these two classes were divided by, among other properties, their Sersic indices with a cut at $n=2.5$, which is close to the median value for our SDSS sample.  The overall conclusion, compared to the case where $M/L$ is assumed to be constant with radius, is unchanged, as both radial regimes undergo mass growth since $z \sim 1$ but with the outer regions building up relatively more mass.

\section{Discussion}

\subsection{A Comparison with the Stellar Mass Growth in the Solar Neighborhood}

Given the significant mass assembly found at large radii, we compare our results in such a region of our galaxy that has been well documented, the solar neighborhood.  Though we caution that the Milky Way is just one such SFG, and may not be one that is an archetypal late-type \citep{hammer2007}, this analysis nevertheless provides an intriguing comparison between our lookback study and galactic archeology.  The solar symbol in Figure~\ref{fig_mass_profiles}(a) indicates the stellar mass surface density at the solar radius \citep[$R_0=8.0$~kpc;][]{vanhollebeke2009} for the Milky Way from \citet{bovy2012b} but scaled down by $0.2$~dex to account for the difference in mass with our descendant SFGs at $z\sim0$.  The systematic uncertainty in this correction may account for the slight offset from the SDSS mass profile, though the rough agreement is still remarkable.  \citet{aumer2009} estimate a mean formation time for stars in the solar neighborhood that corresponds to $z\sim0.9$.  Assuming this mean is close to the median formation redshift, therefore implying a factor of $\sim2$ growth below $z\lesssim0.9$, this estimate is close to the factor of $\sim 2.4\pm0.3$ growth in our mass surface density from $z\sim0.9$ (green line) to the SDSS mass profile at $R=8$~kpc.

\subsection{Caveats}

Two caveats to the analysis presented here warrant some consideration.  First, the contribution of stars formed ex situ to the stellar mass growth is an uncertain quantity, though there is evidence that it is minimal.  Both \citet{behroozi2012} and \citet{moster2013} find that for local halos of mass $M_{\rm h} \sim 10^{12}$~\msun\ (hosting centrals of stellar mass \massmw), little stellar mass was accreted.  At the lowest redshifts, \citet{behroozi2012} find a growing contribution, but this is likely driven by the dominance of QGs in their halos at low redshift, which primarily grow from mergers.  Major mergers are likely rare in our sample at $z\lesssim1$ as such events tend to destroy disks, contrary to what is observed at low redshift (Figure~\ref{fig_mwps}).

Second, the QG fraction increases toward low redshift and as a result not all SFGs at $z\sim 1.3$ will be progenitors of SFGs at low redshift as some will have quenched.  This would impact our results if there is a relation between the structure of high redshift SFG progenitor candidates and low redshift QG descendants, as the former would bias our measurements for the median structural property.

\subsection{Conclusions}

In this paper, we have used the evolution in the SSFR-mass relation to determine the expected stellar mass growth \citep[e.g.,][]{leitner2012} for progenitors of SFGs like the Milky Way.  We used 3D-HST redshifts and photometry to select SFGs of the appropriate progenitor mass at different redshifts back to $z\sim1$ and {\em HST} CANDELS imaging to follow their structural evolution.  As these SFGs grew in stellar mass by a factor of $\sim2$ since $z\sim1$, most of the new stellar mass assembled in the outer regions, leading to a factor of $\sim1.4$ increase in the half-light radius.  The mass surface density profiles indicate ongoing stellar mass assembly also in the central regions where bulges and pseudobulges are common features in spirals.  We also found good agreement between the mass growth at $R=8$~kpc in our lookback study and that for the solar neighborhood of the Milky Way.

\acknowledgments
We thank Daniel Szomoru and Sean McGee for helpful discussions.




\end{document}